\documentclass[aps,prb,twocolumn]{revtex4-1}
\usepackage{times}
\usepackage{graphicx}
\usepackage{amsmath}
\usepackage{bm}
\usepackage{amssymb}
\usepackage{euscript}
\usepackage{verbatim}
\usepackage{fancyhdr}
\usepackage{wrapfig}
\usepackage{setspace}
\usepackage{xcolor}
\usepackage{amsfonts}
\usepackage{subfigure}

\newcommand{\be}{\begin{equation}}
\newcommand{\ee}{\end{equation}}
\newcommand{\bea}{\begin{eqnarray}}
\newcommand{\eea}{\end{eqnarray}}

\newcommand{\mytitle}{Majorana Zero Modes in Synthetic Dimensions}

\begin{document}

\title{\mytitle}

\author{Or Golan}
\affiliation{Raymond and Beverly Sackler School of Physics and Astronomy,
Tel-Aviv University, IL-69978 Tel Aviv, Israel}

\author{Eran Sela}
\affiliation{Raymond and Beverly Sackler School of Physics and Astronomy, Tel-Aviv University, IL-69978 Tel Aviv, Israel}

\author{Kirill Shtengel}
\affiliation{Department of Physics and Astronomy, University of California,
Riverside CA 92511, USA}

\begin{abstract}
Recent experimental advances in the field of cold atoms led to the
development of novel techniques for producing synthetic dimensions and
synthetic magnetic fields, thus greatly expanding the utility of cold atomic
systems for exploring exotic states of matter. In this paper we investigate
the possibility of using experimentally tunable interactions in such systems
to mimic the physics of Majorana chains, currently a subject of intense
research. Crucially to our proposal, the interactions, which are local in
space, appear non-local in the synthetic dimension. We use this fact to
induce coupling between counter-propagating edge modes in the quantum Hall
regime. For the case of attractive interactions in a system composed of two
tunneling-coupled chains, we find a gapless quasi-topological phase with a
doubly-degenerate ground state. While the total number of particles in the
system is kept fixed, this phase is characterized by strong fluctuations of
the pair number in each chain. Each ground state is characterized by the
parity of the total particle number in each chain, similar to Majorana wires.
However, in our system this degeneracy persists for periodic boundary
conditions. For open boundary conditions there is a small splitting of this
degeneracy due to the single-particle hopping at the edges. We show how
subjecting the system to additional synthetic flux or asymmetric potentials
on the two chains can be used to control
this nonlocal qubit.  We propose experimental probes for testing the nonlocal
nature of such a qubit and measuring its state.
\end{abstract}

\maketitle
\section{Introduction}
Physical systems exhibiting topological order, interesting in their own right,
have become a subject of intense attention recently due to their potential
utility for quantum information processing~\cite{Nayak2008}. Much of the recent
experimental effort has been focused on one- and quasi-one-dimensional systems
hosting Majorana zero modes~\cite{Alicea2012a,Beenakker2013a}, in part due the
recent advances in fabricating these systems using semiconducting nanowires,
chains of atoms deposited on the surface of a superconductor and other similar
systems. Meantime, a steady progress in the field of cold atoms led to the
creation a new experimental toolbox~\cite{Goldman2016}, allowing new avenues for
testing similar ideas outside of the realm of condensed matter systems.
Interactions between atoms in cold atomic systems can be custom-tailored by
coupling individual atomic states to light. At the same time, recent advances
led to the dual possibility of creating synthetic gauge fields and endowing
these systems with an extra synthetic dimension~\cite{Celi2014}. Using this
approach, two recent milestone experiments demonstrated realizations of quantum
Hall-like states and their associated chiral edge modes in synthetic ribbons
with artificial gauge fields, one using fermions~\cite{Mancini2015} and the
other one -- bosons~\cite{Stuhl2015}.


In addition, this approach opens interesting new possibilities for quantum
engineering of topological states, e.g. 4D quantum Hall
states~\cite{Price2015}, or for inducing strong correlation effects in low
dimensions, e.g. in magnetic crystals~\cite{Barbarino2015} where one could
observe fractional charge pumping~\cite{Zeng2015,Taddia2017} and probe
signatures of chiral Laughlin-like edge
states~\cite{Petrescu2015,Cornfeld2015,CalvaneseStrinati2017,Petrescu2017}.

In this paper, we discuss the possibility of realizing topological states with
Majorana-like zero modes within the aforementioned approach which relies on
synthetic dimensions/synthetic gauge fields. Specifically, we demonstrate the
appearance of such modes in a cold atomic system where a pairing interaction is
induced between quantum Hall-like states ``separated'' in the synthetic
dimension. Our proposal builds on an earlier proposal for inducing
superconducting proximity in the helical edges of 2D topological
insulators~\cite{Fu2008}, an influential idea which led to subsequent proposals
for Majorana zero modes in semiconductor nanowire
settings~\cite{Lutchyn2010a,Oreg2010} as well as more recent fractional
generalizations~\cite{Clarke2013a,Lindner2012,Cheng2012,Alicea2016}. Crucially,
our setup is different from the usual condensed matter schemes in that it
utilizes a closed system with particle
conservation~\cite{Cheng2011,Sau2011,Kraus2013,Ruhman2015,Iemini2015,Mazza2015,Chen2017,Guther2017}.

\begin{figure*}[ht]
	\centering	
	\includegraphics[scale=0.253]{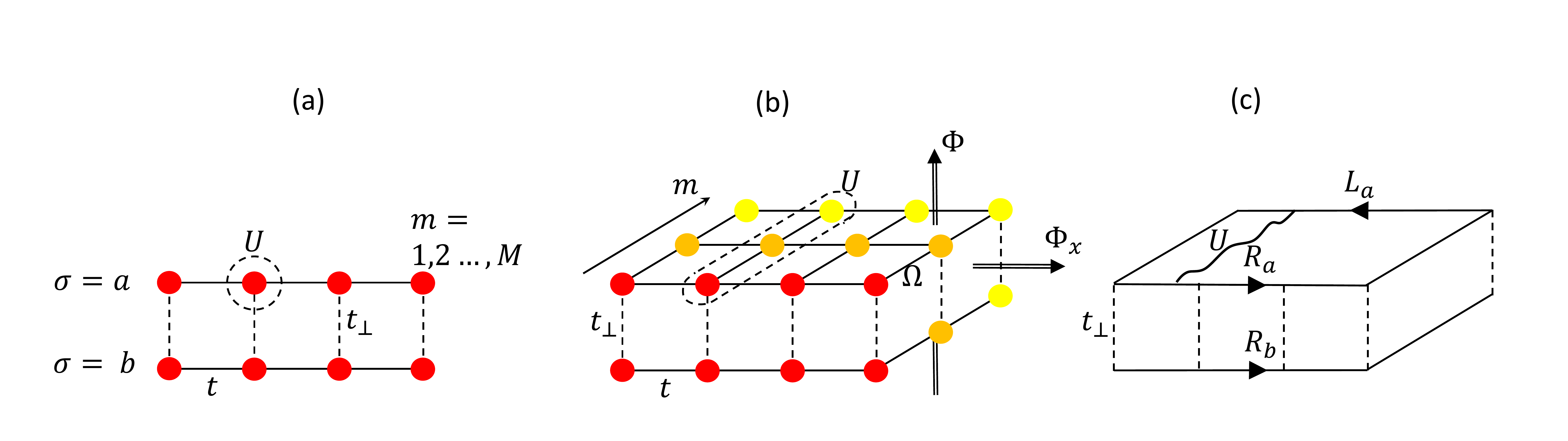}
	\caption{(a) Two coupled chains with an on-site interaction $U$ and internal
atomic quantum number $m=1,...,M$. $m$ is conserved upon either intrachain ($t$)
or interchain ($t_\perp$) hopping. (b) Equivalent picture where quantum number
$m$, $1 \le m \le M$, is interpreted as an additional ``dimension".
In addition a finite transition amplitude $m \to m+1$ is included ($\Omega$)
with imprinted phase $\Phi$. (c) In the quantum Hall regime the bulk modes $m=1,M$
are gapped and at low energy the system consists of edge modes.
Counter propagating edge modes interact non-locally via the initially
on-site Hubbard interaction $U$.}
\label{fig:main1}
\end{figure*}

The idea of additional ``synthetic'' dimensions is schematically illustrated in
Fig.~\ref{fig:main1}, with the role of an extra ``dimension'' played by an
internal atomic degree of freedom, such as nuclear spin. This extra dimension
is intrinsically both discrete and finite, nonetheless it allows one to
effectively turn a physical 1D atomic chain into a 2D strip/ladder.

A key feature of the synthetic dimension approach is that interactions become
\emph{non-local} in the synthetic dimension~\cite{Celi2014}; see
Fig.~\ref{fig:main1}(a,b). This opens a new possibility, which we exploit in
our proposal: namely, it allows coupling between the degrees of freedom which
are normally spatially separated in the usual condensed matter setting. In
particular, this enables us to create attractive interactions between
counter-propagating quantum Hall edge states~\cite{Yan2015} in order to induce
superconducting instabilities; see Fig.~\ref{fig:main1}(c).

In what follows, we will focus on a system consisting of two identical chains,
forming two synthetic ribbons in the quantum Hall regime. Using the
renormalization group analysis~\cite{Cheng2011}, we show that this closed
system forms a many-body phase with strong pair-tunneling coupling between the
two chains. Its ground state is doubly degenerate ground state, resembling that
of the Kitaev chain~\cite{Kitaev2001}. However, the total number of particles
is fixed in our case. This is the crucial difference between our approach and a
related earlier proposal presented in Ref.~[\onlinecite{Yan2015}] where a
single Hall ribbon has been treated in a BCS mean-field approximation. By its
nature, such a mean-field approximation breaks particle number conservation,
leading the authors of Ref.~[\onlinecite{Yan2015}] to a physically tenuous
conclusion about the existence of Majorana zero modes in their setup. In
contrast, our approach \emph{does not} rely on the mean-field approximation; we
show that the prerequisite pairing instability is triggered by arbitrarily weak
attractive interactions. Nevertheless, the presence of interactions is crucial
here; zero modes found  in a related, but non-interacting setup in
Ref.~[\onlinecite{Klinovaja2013a}] are of the
Su--Schriefer--Heeger~\cite{Su1980}, not Majorana type. (In particular, those
zero modes can be individually occupied or empty, implying e.g.\ the wrong
quantum dimension that is inconsistent with the claimed non-Abelian braiding
statistics of the Ising type.) Meantime the particle conservation constraint is
circumvented in our case by considering a double synthetic ribbon. The ground
state degeneracy is no longer associated with the overall fermion parity of the
closed system and is instead encoded in the parity in each chain.

The crucial advantage of using the quantum Hall regime is that it naturally
allows for generalizations to the fractional case, where we expect that with
small modifications the present setup will allow an experimental realization of
fractional topological superconductor phases containing exotic anyons, e.g. of
the parafermion type. In this paper we shall concentrate on the proof of
concept for the simplest possible case, leaving such generalization to
fractional state to the future.

\section{Model}
We consider a double chain, or two-leg ladder, of atoms with an internal
quantum number $m=1,..,M$, as shown in Fig.~\ref{fig:main1}(a).  Atoms can hop
along the chain or between the two chains with hopping amplitudes $t$ and
$t_\perp$ respectively. These hopping processes conserve the internal quantum
number $m$. In addition, our model allows internal transitions $m \to m \pm 1$
with amplitude $\Omega$ and an imprinted phase $\Phi$, which can be achieved in
practice by illuminating the system by additional lasers at judiciously chosen
angles~\cite{Celi2014}. These transitions can be regarded as hopping in the
transverse `synthetic' direction. I.e., each leg of the physical ladder is now
effectively a strip, as shown in Figure~\ref{fig:main1}(b). Finally, an on-site
interaction $U$ -- see in Fig.~\ref{fig:main1}(a) -- becomes in effect a
non-local interaction within each synthetic strip. The resulting tight binding
Hamiltonian for our system is $H_\text{lattice}= H_{t} + H_\perp + H_\Omega +
H_{int}$ where
\begin{subequations}
 \label{eq:lattice}
\begin{align}
&H_{t} =-t \sum_{x,\sigma,m} ( c^\dagger_{x,m,\sigma}c_{x+1,m,\sigma} + h.c. ), \label{eq:lattice_a}\\
&H_\perp = t_\perp\sum_{x,m}\left( e^{i  \Phi_x \frac{m}{M}} c^\dagger_{x,m,a}c_{x,m,b} + h.c. \right), \label{eq:lattice_b}\\
&H_\Omega = \Omega \sum\limits_{x,\sigma} \sum\limits_{m=1}^{M-1}
    \left(e^{ix\Phi}c^\dagger_{x,m,\sigma}c_{x,m+1,\sigma} + h.c. \right), \label{eq:lattice_c}\\
&H_{int} =U \sum_{x,\sigma}  n_{x,\sigma}^2. \label{eq:lattice_d}
\end{align}
\end{subequations}
Here, $\sigma = \{ a,b\}$ labels the legs (strips) of the ladder, and $n_{x,
\sigma} = \sum_{m}c^\dagger_{x,m,\sigma}c_{x,m,\sigma}$. The inter-chain
tunneling term~(\ref{eq:lattice_b})  also also allows for a position-dependent
tunneling phase $\Phi_x$, which will be discussed later.

In what follows, we focus on the case of strongly anisotropic tunneling,
specifically $t_\perp \ll t$. We envision each of the synthetic strips to be in
the quantum Hall regime, so that the system forms two weakly coupled quantum
Hall strips, which requires the following relation between the model
parameters: \bea |U|, t_\perp \ll \Omega \ll t. \eea 

\begin{figure}[ht]
	\centering	
	\hspace*{-.25in}
    	\includegraphics[scale=0.353]{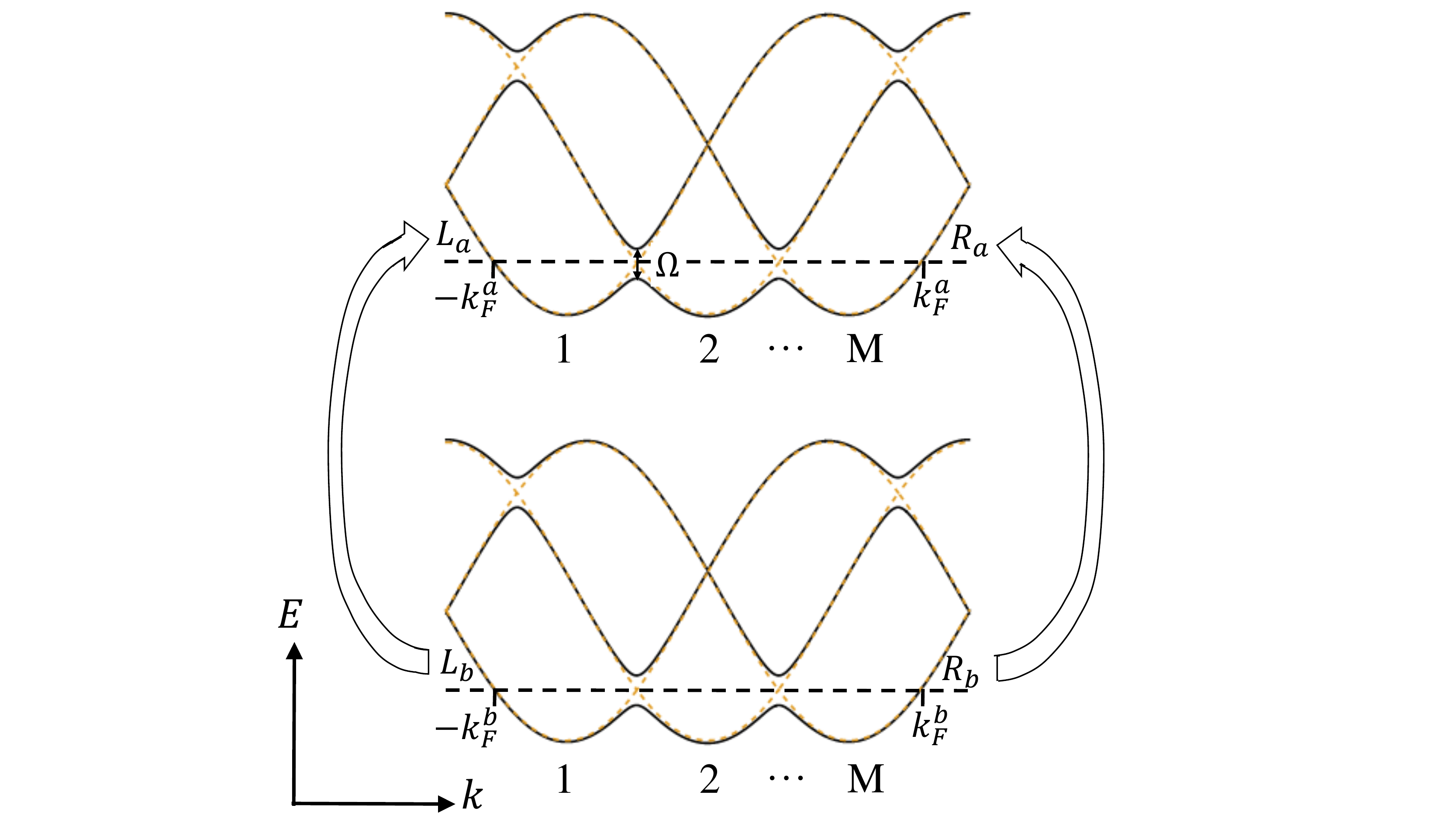}
	\caption{Energy-momentum diagram and interaction processes of the model Eq.~(\ref{eq:lattice}).
The cosine dispersions for each $m$ are shifted in momentum space due to the flux $\Phi$ (dashed lines).
The $m \to m \pm 1$ processes $\Omega$ create avoided crossings and open energy gaps separating
Landau levels (full lines). Two counter propagating edge modes $R_\sigma$, $L_\sigma$
remain at the Fermi level (dashed line) for each chain $\sigma = a,b$. Among interaction processes,
the pair tunneling term transfers two particles between the two chains.
It is facilitated by the non-local interaction $U$ between the edge states.
}
\label{fig:main2}
\end{figure}

Let us first consider the case of decoupled strips, $t_\perp=0$. A single
particle dispersion relation in the absence of the interaction
term~(\ref{eq:lattice_d}) is shown in Fig.~\ref{fig:main2}.  When there is no
tunneling in the synthetic dimension as well ($\Omega=0$), the dispersion
relation for each strip ($\sigma=a,b$) consists of $M$ cosines (dashed lines in in
Fig.~\ref{fig:main2}), shifted with
respect to one another horizontally in the momentum space due to the presence
of the synthetic magnetic flux $\Phi$. For finite $\Omega$, $m \to m \pm 1$ transitions lead to
avoided crossings and open gaps (solid lines in Fig.~\ref{fig:main2}). Each
synthetic strip thus realises a coupled-wire construction of
Ref.~[\onlinecite{Kane2002}]. We assume that the temperature is lower than the
quantum Hall gap $\Omega$, allowing one to reach the quantum Hall regime.
Specifically we consider the filling factor $\nu=1$, meaning that the Fermi
level lies within the first gap, as shown in Fig.~\ref{fig:main2}. This corresponds
to the number of atoms per (synthetic) site being $\langle n_{x,m,\sigma}
\rangle = \Phi / 2 \pi$.

In the quantum Hall regime all modes in the bulk ($1 < m < M$) are gapped and
the low energy physics is governed solely by the two chiral edge states crossing the
Fermi level. Specifically, for each physical chain of atoms $\sigma=a,b$, the
first synthetic chain $m=1$ hosts a left-moving mode, and the $M$-th synthetic
chain hosts a right-moving edge mode, \bea \label{cRL} c_{x,M,\sigma} = e^{i
k_F^\sigma x} R_\sigma(x),~~~c_{x,1,\sigma} =  e^{-i k_F^\sigma x} L_\sigma(x).
\eea Here $R_\sigma(x)$, $L_\sigma(x)$ are slowly varying fermionic fields (we
have set the short distance cut-off $a_0=1$). The effective edge Hamiltonian is
$H_{edge} =\int dx \left( \mathcal{H}_0 + \mathcal{H}_\perp \right)$, where 
\begin{subequations}
\label{fermionmodel}
\begin{align}
\label{fermionmodel_a}
\mathcal{H}_0 &= \sum_\sigma  \left(i v_\sigma
\left[R_\sigma^\dagger \partial_x R_\sigma
-L^\dagger_\sigma \partial_x L_\sigma\right]+  U   R^\dagger_\sigma R_\sigma L^\dagger_\sigma L_\sigma \right) \\
\label{fermionmodel_b}
\mathcal{H}_\perp &=  t_{\perp}  \left(e^{ -i \delta k x } e^{i\Phi_x/2}R^\dagger_{a}R_b +e^{ i \delta k x } e^{-i\Phi_x/2}L^\dagger_{a}L_{b}\right) + h.c.
\end{align}
\end{subequations} The first term contains both the kinetic energy and density--density
interaction within each physical chain. Here $v_\sigma = 2 t \sin(k_F^\sigma)$
where $k_F^\sigma=n_\sigma \pi$ with $n_\sigma = \langle \hat{n}_{x,\sigma,m}
\rangle$ being the average density in each synthetic chain. We shall ignore
band curvature and hence assume $v_a=v_b=v_F$.
The second term describes tunneling between the two physical chains; using
Eq.~\eqref{cRL} we can see that for a finite density difference between the two
chains this term oscillates as $e^{i \delta k x}$ where $\delta k =
k_F^a-k_F^b$ and is therefore irrelevant as long as $k_F^a \ne k_F^b$.

\section{Superconducting phase for attractive interactions}
With the goal of realising topological superconducting states, we now turn to
the analysis of our model in the case of attractive interactions, $U<0$, which
is a realistic possibility in the realm of cold atoms. 	For our analysis to be
self-contained, we shall review some previous results on Majorana fermions in
setups with particle number conservation~\cite{Cheng2011,Keselman2015,
Chen2017}, whose effective description closely matches that of
Eq.~(\ref{fermionmodel}). We will emphasize features and experimental knobs
that are specific to our proposed setup based on synthetic quantum Hall
ribbons, particularly the role of the non-locality of interactions in the
synthetic dimension and the flux $\Phi_x$ inserted within a loop in the plane
containing the synthetic dimension. We begin by characterizing the resulting
superconducting phase from the point of view of a Luttinger liquid instability
to pair tunneling, see Fig.~\ref{fig:main2}, by following the analysis and
notations of Cheng and Hao~\cite{Cheng2011} who studied an identical effective
model. We will address its ground state degeneracy using
bosonization. In close analogy to the analysis of Ref.~\onlinecite{Chen2017},
we will construct Majorana operators and show that the model has an emergent
$Z_2$ symmetry obtained at a special value of the flux $\Phi_x$; when projected
to the low energy subspace, this $Z_2$ symmetry coincides with the parity
symmetry generated by the Majorana operators and hence protects the ground
state degeneracy from local perturbations.

\subsection{Renormalization group analysis}
In the presence of interactions, the inter-chain tunneling [see
Eq.~(\ref{fermionmodel_b})] results in the generation of two more quartic terms
in the effective low-energy Hamiltonian: simultaneous backscattering within the two
physical chains, and pair tunneling  between them. These terms are proportional
to $\left(e^{2 i \delta k x} L_a^\dagger R_a R_b^\dagger L_b + h.c.\right)$ and
$R_a^\dagger L_a^\dagger L_b R_b$ respectively.
As can be seen from Fig.~\ref{fig:main2}, at energies much smaller than
$\Omega$ these processes involve just the edge modes. Under normal
circumstances the amplitudes of these processes would be suppressed due to the
spatial separation of the counter-propagating edge modes. However, in our
setting this separation occurs in the synthetic dimension, resulting in no such
suppression: the interaction $U$ is nonlocal in this dimension.

Notice that for a finite density imbalance between the chains, $\delta k =
k_F^a - k_F^b \ne 0$, while both the single particle tunneling described by
Eq.~(\ref{fermionmodel_b}) as well as the aforementioned simultaneous
backscattering operator do not conserve momentum and, as a result, oscillate
with wavenumbers $\delta k $ and $2 \delta k$ respectively. Meantime, the
induced pair tunneling term does not oscillate. This property can be used to
selectively promote the pair tunneling process.

It is convenient to bosonize this model,
\begin{equation}
R(L)_\sigma = \frac{1}{\sqrt{2\pi a_0 }} e^{i\sqrt{\pi}(\theta_\sigma+r\varphi_\sigma)},
\end{equation}
where $r=+/-$ for $R/L$ and $\sigma=a,b$; $a_0$ is a short distance cutoff, and
the two bosonic fields satisfy $[\partial_x \varphi_\sigma (x), \theta_\sigma
(x')]= i \delta (x-x')$. The field $\varphi_\sigma$ is related to  the charge
density in chain $\sigma$ via $\rho_\sigma = \frac{1}{\sqrt{\pi}} \partial_x
\varphi_\sigma (x)$, and its conjugate field $\theta_\sigma$ may be interpreted
as the phase of the pair field $L_\sigma R_\sigma$. Finally we define even and
odd combinations $\varphi_\pm = \frac{1}{\sqrt{2}}(\varphi_a \pm \varphi_b)$
and $\theta_\pm = \frac{1}{\sqrt{2}}(\theta_a \pm \theta_b)$.

In the bosonized language the edge Hamiltonian $H_\text{edge}$, which includes
pair tunneling and backscattering, becomes $ H = H_0 + H_\perp + H_2$ where
\begin{subequations}
\label{bosonizedH}
\begin{align}\label{bosonizedH_a}
&H_0 =\sum_{\mu = \pm} \frac{v_\mu}{2}\left[ K_\mu(\partial_x\theta_\mu)^2 + K_\mu^{-1}(\partial_x\varphi_\mu)^2\right],  \\
\label{bosonizedH_b}
&H_\perp = \frac{2t_\perp}{\pi a_0}\cos\!\left(\sqrt{2\pi}\tilde\varphi_- - {\Phi_x}/{2}\right)\cos\sqrt{2\pi}\theta_- ,  \\
\label{bosonizedH_c}
&H_{2} = \frac{g_1}{(\pi a_0)^2}\cos\sqrt{8\pi}\tilde\varphi_- +
\frac{g_2}{(\pi a_0)^2}\cos\sqrt{8\pi}\theta_- ,
\end{align}
\end{subequations}
with $K_+ = K_- = 1- O(U)$ being the Luttinger parameter, $v_+ = v_-$
being the Luttinger liquid velocity, and $g_1, g_2 \propto t_\perp^2$. In the
last two lines $\tilde\varphi_{-}\equiv \varphi_{-}+\delta k x/\sqrt{2\pi}$,
which accounts for the density imbalance between the two chains.


This form of the Hamiltonian allows us to discuss instabilities of the
Luttinger liquid. Notice that the even sector ($\mu=+$) remains gapless.
The odd sector ($\mu=-$) contains both the single particle tunneling
[Eq.~(\ref{bosonizedH_b})] and the two-particle processes
[Eq.~(\ref{bosonizedH_c})].
Following the RG analysis of Ref.~\onlinecite{Cheng2011}, we find that the
single particle tunneling has scaling dimension $x_\perp =(K_-+K_-^{-1})/2$ and
should therefore should be relevant ($x_\perp<2$) even for  $U=0$.
However, for a finite density mismatch between the chains (which we assume
here), it becomes oscillating and hence irrelevant.
The correlated backscattering and pair tunneling terms have scaling dimensions
$x_b=2K_-$ and $x_p=2K^{-1}_-$ respectively; for arbitrarily small attraction,
$U<0$, we have $K_->1$ so that pair tunneling becomes relevant ($x_p<2$) while
backscattering -- irrelevant ($x_b>2$).

Hence even weak attractive interactions in the presence of density imbalance
between the chains would stabilize a quantum phase dominated by pair tunneling.
We now turn to a discussion of the properties of this phase.
\subsection{Ground state degeneracy}
We consider a system of linear extent $L$ (i.e. $0< x < L$), whose behavior is
dominated by the relevant pair tunneling term $\cos\sqrt{8\pi}\theta_-$. It
locks the $\theta_-$ field, {i.e.}, the difference in the phases of the pair
fields $L_a R_a$ and $L_b R_b$, to $\theta_-=\sqrt{\pi/2} ~n_\theta$ with
integer-valued operator $n_\theta$. 	

Generally, the ground state degeneracy associated with a topological nontrivial
phase depends on the boundary conditions.  Our 1D system dominated by pair
tunneling is surrounded by trivial regions both on the left ($x<0$) and the
right ($x>L$). We can mimic these boundaries by introducing a large mass term $
M\sum_{\sigma} \int dx R^\dagger_\sigma L_\sigma+h.c.$ to the trivial
regions~\cite{Keselman2015}. Equivalently, after bosonization, this term can be
written as
\begin{multline}\label{cosphi}
H_{M}= \sum_{\sigma} \int \frac{M}{ \pi a_0} \cos(2 \sqrt{\pi} \varphi_\sigma) \\
= \int \frac{2M}{ \pi a_0} \cos( \sqrt{2 \pi} \varphi_+ )\cos(  \sqrt{2\pi}
\varphi_-).
\end{multline}
In the strong coupling limit, it pins the $\varphi_\pm$ fields on the left and
right sides, $\varphi_-(x) = \sqrt{2 \pi} n_\varphi^L$, $(x<0)$ and
$\varphi_-(x) = \sqrt{2 \pi} n_\varphi^R$, $(x>L)$, with $n_\varphi^{L/R}$
being integer-valued operators.~\footnote{For finite density mismatch $\delta k
\ne 0$ the integer-valued operators $n_\varphi$ are defined in the same way in
terms of $\tilde{\varphi}_-$ rather than $\varphi_-$.}

 These integer eigenvalues have a transparent
physical meaning. First consider the individual chains $\sigma = a,b$. The
difference $\varphi_\sigma(x>L)-\varphi_\sigma(x<0) = \sqrt{\pi} n_\sigma$
according to Eq.~(\ref{cosphi}). Using $\rho_\sigma = \frac{1}{\sqrt{\pi}}
\partial_x \varphi_\sigma$, we see that $n_\sigma$ is the number of particles
in chain $\sigma$. Conservation of the total particle number constraints
$n_a+n_b=\text{const}$. Let us focus on the case when this number is even.
Since both $n_{a}$ and $n_b$ are integer, the difference $n_a-n_b $ must be
even as well.
The individual particle numbers $n_a$ and $n_b$ are not conserved by the
Hamiltonian due to both the single particle and pair tunneling terms.
The pair tunneling operator changes this number by 2 in each chain, thus
preserving their individual parities. Hence it commutes with parity
$\hat{\mathcal{P}}=e^{i \pi \delta Q}$, where $\delta Q = {\left(n_a - n_b\right)}/{2} =
n_\varphi^R - n_\varphi^L$.



Therefore, if we ignore the single particle tunneling term (we shall return to
this point later), the entire low energy spectrum (and not just the ground state!) is
two-fold degenerate: the degeneracy corresponds to $\hat{\mathcal{P}}=\pm 1$.
We denote the corresponding states as $\vert e \rangle$ and $\vert o \rangle$
respectively.

In the bulk, $0<x<L$, the pair tunneling operator pins the $\theta_-$ field to
$\theta_-= \sqrt{\frac{\pi}{2}} n_\theta$ with integer-valued $n_\theta$. Thus
one can consider the operator $e^{i \sqrt{2 \pi} \theta_-} \equiv e^{i \pi
n_\theta}$, and try to distinguish states within each parity sector, with
different locking of $\theta_-$. However, despite of this additional pinning,
the states $|e \rangle$ and $|o \rangle$ with well defined $\delta Q=n_\varphi^R -
n_\varphi^L$ can not be further distinguished by the value of $\theta$ since
the integer-valued operators $n_\varphi^{R/L}$ and $n_\theta$ do not all
commute. Using the non-local commutation relation between $\varphi_{-}(x)$ and
$\theta_{-}(x)$, and projecting them to the low energy subspace, one arrives at
\begin{equation}\label{eq:commutation}
\left[n^R_\varphi,n_\theta\right] = \frac{i}{\pi}, \qquad \left[n^L_\varphi,n_\theta\right] =0.
\end{equation}
This implies that $\hat{\mathcal{P}}=e^{i \pi \delta Q}$ and $\hat{\Theta}=e^{i
\pi n_\theta}$ anti-commute rather than commute, $\hat{\mathcal{P}}
\hat{\Theta} =- \hat{\Theta} \hat{\mathcal{P}}$. Consequently, working in the
$\hat{\mathcal{P}}$ basis one finds that $\langle {\mathcal{P}}|e^{i \pi
n_\theta} |{\mathcal{P}} \rangle=0$. Therefore, the low-energy spectrum of the
system is indeed only two-fold degenerate.

The above algebra suggests a Pauli matrix representation of the aforementioned
operators, \be \hat{\mathcal{P}}= \sigma^z,~~~\hat{\Theta} = \sigma^x. \ee One
can change basis from the parity basis to the eigenbasis of $\hat{\Theta}$, as
$|{\Theta} \rangle = \frac{1}{\sqrt{2}} \left(|e \rangle \pm  |o \rangle
\right)$.


\subsection{Majorana operators, single particle tunneling, and $Z_2$ symmetry}
\label{se:z2}
Following Chen et. al.~\cite{Chen2017} it is natural to define Majorana operators
	\be
	\label{MAJORANA}
	\gamma_{L/R} = e^{i \pi (n_\varphi^{L/R}+n_\theta)},
	\ee 
which are Hermitian, square to one, and anti-commute. In terms of these
operators, the parity operator $\hat{\mathcal{P}} = i \gamma_L \gamma_R$. We
note, however, that no symmetry protects the ground state degeneracy associated
with this parity in the presence of a finite single particle hopping term,
$t_\perp\neq 0$. Even if irrelevant, it can be effective near the edges, i.e.,
a local perturbation such as $H_\perp$ near the left or right edge can couple
to one of the Majorana operators and change the parity. However, this coupling
can be eliminated by tuning flux $\Phi_x$ to a specific value. A special
symmetry emerges in this case and prohibits coupling to the individual Majorana
operators.


Consider the bosonized Hamiltonian given by Eq.~(\ref{bosonizedH}) at
$\protect{\Phi_x=\pi}$. One can identify the special symmetry \bea
U:~ \theta_- & \to & \theta_- + \sqrt{\pi/2}, \nonumber\\
  \tilde{\varphi}_- & \to &  -\tilde{\varphi}_-  .
\eea The latter transformation implies ${\varphi}_-  \to  - {\varphi}_-$ and
$x\to -x$. The transformation $\theta_-  \to \theta_- + \sqrt{\pi/2}$ shifts
$\theta_-$ between subsequent minima of the pair-tunneling term, but changes
the sign of $\cos ( \sqrt{2 \pi} \theta_ -)$ in the single particle hopping
term, which however gets compensated by $ \tilde{\varphi}_ - \to  -
\tilde{\varphi}_-$ exactly at $\Phi_x = \pi$.

In addition, one can see how this transformation acts on the Majorana
operators.  Since it shifts $n_\theta$ by unity, and takes $n_\phi \to - n_\phi
= n_\phi~ {\text{mod}}~ 2$, we may conclude that this symmetry $U$ acts on the
Majorana fermions as \be U \gamma_{L/R} U^{-1}=  - \gamma_{L/R}. \ee Thus
within the low-energy subspace we can identify $U$ with the parity operator,
\be U = \hat{\mathcal{P}}\equiv i \gamma_L \gamma_R , \ee which acts in the
same way on the Majorana operators. Therefore, by fine tuning the parameter
$\Phi_x$ we can reach a point  where a coupling to a single Majorana fermion,
which could change the parity quantum number $P$, becomes forbidden by
symmetry. We shall see this mechanism in action in the calculation of matrix
elements of the single particle tunneling term presented in the next section.
We note that the symmetry discussed here does not correspond to any microscopic
symmetry. For instance, it is  different from the time reversal symmetry
discussed in a similar context in~[\onlinecite{Haim2016}].

It is instructive to bosonize the single particle tunneling operator, replace
$\varphi_-$ and $\theta_-$ by their expressions in terms of integer-valued
operators, and compare the resulting expression with that of the Majorana
operators $\gamma^{R/L}$ in Eq.~(\ref{MAJORANA}):
\be
L^\dagger_a L_b (x=0) \sim e^{i \pi (2 n_\varphi^{L/R}+n_\theta)}.
\ee
Note that the single Majorana operators in our strongly interacting state are
nonlocal in terms of the original particles.

 \subsection{Finite splitting for $\Phi_x \ne \pi$}
We shall now address the single-particle tunneling process $H_\perp$ near the
edges (i.e., the zero-dimensional boundary) in more detail, focusing on its
effect on the approximate two-fold degeneracy. Although no local operator can
measure the total parity and hence distinguish the $ | e\rangle$ and $ | o
\rangle$ states, it is possible to mix these states by a local process --
tunneling of a single particle.

Let us express the single particle tunneling operator of
Eq.~(\ref{bosonizedH_b})
near the left edge, $x=0$, or the right edge, $x=L$, in terms of Pauli matrices
$\sigma^i$. $\mathcal{H}_\perp(x)$ is off-diagonal in the parity basis since it
transfers one particle between the chains, and therefore it has to be a
combination of $\sigma^x$ and $\sigma^y$. In addition, as it contains $e^{2 \pi
i n^{L/R}_\varphi }$ rather than $e^{\pi i n^{L/R}_\varphi }$, it is diagonal
in the $|\Theta \rangle$ basis. Thus $\mathcal{H}_\perp(x)$ acts in the ground
state manifold as $\sigma^x$. To compute its $x$ dependence near the interfaces
we approximate the cosine potential by a mass term in the appendix, and obtain
\be \label{hperp} \langle \Theta |\mathcal{H}_\perp (x) | \Theta \rangle =
\frac{2 t_\perp}{\pi \xi} \cos (\Phi_x/2)   e^{i \pi n_\theta } e^{-\pi x/(2
\xi)}, \ee with correlation length $\xi =  v / \Delta$ and    energy gap
$\Delta \propto (g_2)^{\frac{1}{2-x_p}}$ (here we have assumed $K_-=1$; for the
dependence on $K_-$ see the appendix). Integrating over $x$ one obtains \bea
\langle \Theta |H_\perp |\Theta \rangle = \frac{4 t_\perp}{\pi^2} \cos
(\Phi_x/2)   e^{i \pi n_\theta }. \eea

The exponential decay of these matrix elements could be tested numerically by
computing the $x$ dependence of the operator the operator $H_\perp(x) =
\sum_{m} c^\dagger_{x,m,1} c_{x,m,2}+h.c.$. In addition, such inter-site
hopping process may be directly measurable in experiment using the techniques
for detecting particle currents~\cite{Atala2014}.

In full agreement with the symmetry argument of the previous section,
the synthetic magnetic flux $\Phi_x$ between the two synthetic ladders
can completely cancel the effect of single particle tunneling within the low
energy subspace. This cancellation has a simple physical interpretation in
terms of a two-path interference. The total result for the matrix element of
$H_\perp$ in Eq.~(\ref{hperp}) is in fact a sum of two independent hopping
terms which turn out to have equal weight: $\langle n_\theta |R^\dagger_{a}R_b
| n_\theta \rangle = \langle n_\theta |L^\dagger_{a}L_b | n_\theta \rangle $ -
one due to the hopping between the right moving modes at one end in the
synthetic dimension, and the other due to hopping between the left moving modes
at the other end. In the presence of the flux $\Phi_x$, the phase difference
between the two paths can result in a destructive interference. This once again
emphasizes the fact that the Majorana operators are non-local in the synthetic
dimension.
\section{Fusion of Majorana fermions and non-local entanglement}
So far we have been considering a single region where superconductivity emerges
intrinsically and leads to an approximate two-fold degeneracy associated with
Majorana edge operators. We now turn to the setting with multiple
superconducting regions separated by trivial regions. Specifically, we will
focus on the process of nucleating such an extra trivial region within a
superconducting region in \emph{real time}, which is a possibility in cold atom
systems.

\begin{figure*}[ht]
	\centering	
	\includegraphics[scale=0.25]{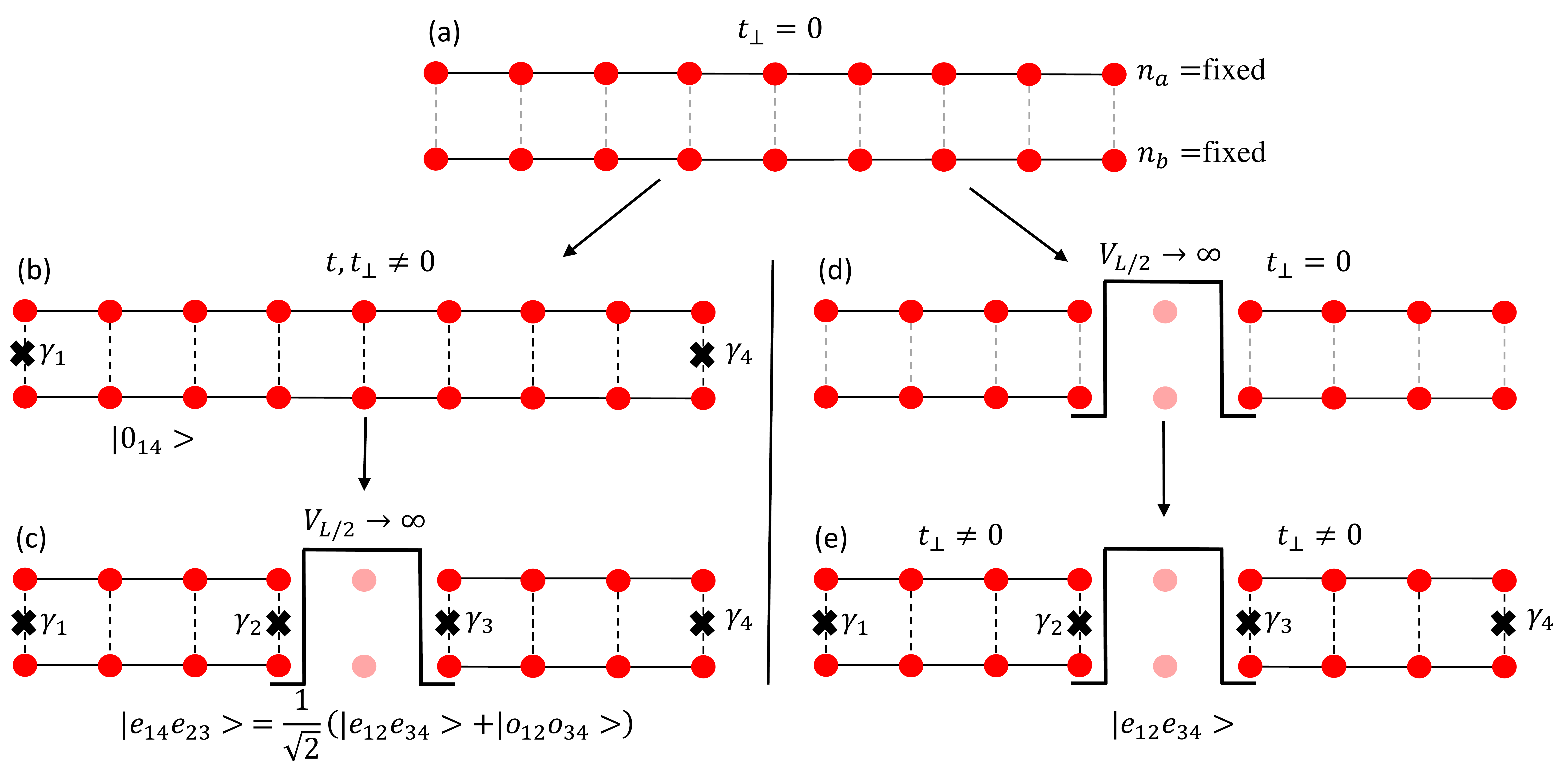}
	\caption{Implementation of a fusion-testing protocol analogous to Ref.~[\onlinecite{Aasen2016}]
    in a cold atom setting with particle number conservation.
    (a) Two disconnected chains with $t_\perp=0$ and a fixed number of particles in each chain $n_a,n_b$.
    (b) By turning on the inter-chain coupling $t_\perp$ the
		system is driven to the doubly degenerate state with well defined parity
		state of Majorana operators $i \gamma_1 \gamma_4 = +1$ dented $| e_{14} \rangle$.
    (c) A strong potential barrier is created in the center. The resulting central
		region has a trivial gap, similar to the
		regions at $x<0$ and $x>L$. Thus an additional degeneracy emerges described by
		two additional Majorana operators $\gamma_2 , \gamma_3$ which have been created
		from the vacuum in an even parity state $|
		e_{23} \rangle$. As described in Eq.~(\ref{fusion}) this is a maximally
        entangled state in terms of the parity of the left and right sides.
        (d,e) To test that parity fluctuations in each subregion are intrinsic to the system,
        we can prepare a system described by the same four Majorana operators
        but with a well defined parity in each half, by first
        breaking the decoupled chains into two separate segments each (d), and then
        turning on the inter-chain coupling $t_\perp$ (e).  \label{fig:fusion}}
\end{figure*}

By analogy with the setup of Ref.~[\onlinecite{Aasen2016}], consider the
process whereby a uniform system is initialized in a well-defined parity state
and then divided it into two separate parts by {e.g.} ramping up the potential
energy term in the central region, as depicted by the step (b)$\to$(c) in
Fig.~\ref{fig:fusion}. This process results in creating a pair of new Majorana
operators, $\gamma_2$ and $\gamma_3$. Consequently, each new superconducting
region can be either in an even or odd parity state, described by the
eigenvalues of the operators $i \gamma_1 \gamma_2$ and  $i \gamma_3 \gamma_4$.
We should remind the reader that, owing to the particle conservation constraint
in each segment, the notion of parity in our case is different from that of
Ref.~[\onlinecite{Aasen2016}]; for us the overall parity of each segment is
fixed while the degeneracy is associated with the individual parity of its two
constituent chains. Let us assume for concreteness that the system is
initialized in an even parity state $i\gamma_{1}\gamma_{4}=1$, denoted as $|
e_{14}\rangle$. Upon adiabatic creation of the middle barrier, two new Majorana
operators$\gamma_{2,3}$ are created from the vacuum and hence are also in an
even parity state $| e_{23}\rangle$. By changing to the basis associated with
the fusion outcomes of Majorana pairs in the right and left halves, i.e.
$\{\gamma_{1,2} \}$  and $\{\gamma_{3,4} \}$, one obtains \be \label{fusion}
 | e_{14} e_{23} \rangle = \frac{1}{\sqrt{2}} \left(| e_{12} e_{34} \rangle+| o_{12} o_{34} \rangle\right).
\ee The parities associated with the of the left and right sides are maximally
entangled. This is a topological effect, as it does not depend on the details
such as the precise asymmetry of the bipartition of the system.

The resulting state contains maximal parity fluctuations in each segment.
Detecting them can be done first preparing the system in the state with no
overall parity fluctuations and then implementing a protocol similar to that of
Ref.~\onlinecite{Aasen2016} and depicted in  Fig.~\ref{fig:fusion}. This test
requires measuring parity states of each segment of the system, using, for
example, time of flight measurements as discussed e.g. in
Ref.~[\onlinecite{Kraus2012}];

The validity of the low-energy effective description of our system in terms of
Majorana zero modes requires adiabaticity of the time dependent process, with
respect to the relevant energy gaps. We should recall, however, that our system
is in fact gapless due to the extra even sector associated with the total
charge. Its effective description is the Luttinger liquid with fields
$\theta_+, \varphi_+$. As discussed in the previous sections and earlier
works~\cite{Cheng2011,Keselman2015, Chen2017}, the topological properties in
the odd sector (i.e. the spin sector), remain approximately impervious to the
charge sector, which therefore acts merely as a spectator. However, as we shall
see, the charge sector plays a crucial role in determining the adiabatic
condition for our ``fusion'' protocol.

While the total charge $N$ is fixed, as the barrier is raised, the total number
of particles in the left side $N_L$ or in the right side $N_R = N-N_L$ can
fluctuate. Yet, it is precisely the Luttinger liquid of the charge sector which
provides a finite charge compressibility. Assuming for concreteness that $N$ is
even, as well as a symmetric left-right bipartition, the lowest energy charge
state will have $(N_L,N_R)=(N/2,N/2)$. The excited charge states
$(N_L,N_R)=(N/2 \pm 1,N/2 \mp 1)$ will have an energy cost of order
$\frac{\hbar v}{L}$ originating from the Luttinger liquid Hamiltonian of the
charge sector. This energy scale sets the adiabatic condition implying that the
system size should not be too large.

As an alternative to this very restrictive adiabatic condition one may in fact
choose the opposite, namely perform a sudden quench of the barrier. After a
sudden ramp-up of the potential energy term in the central region, we end up
with the vacuum state described by Eq.~(\ref{cosphi}). The field $\varphi_+$ is
pinned in the central region, but can take different values $\varphi_+ (L/2) =
\sqrt{\pi/2} N_L$ with $N_L$ being an integer-valued operator describing the
total (summed over $\sigma = a,b$) number of particles in the left side.
Imagine now performing a strong charge measurement of $N_L$. After such a
measurement the quantum state collapses onto one with a fixed $N_L$.
Independently, the value of $\varphi_-$ in the barrier region is described by
the integer valued operator $n_\varphi^\text{barrier}$, taking two physically
distinct values corresponding to the relative parity of the chains in the left
side. The latter is linked with the parity in the right side: for even $N_{L}$
both parities are equal, and for odd $N_{L}$ the two parities are opposite.
Since before the sudden quench the wave function in the central region was
dominated by pair tunneling, with pinned $\theta_-$ field, it follows that this
state is a superposition in terms of $n_\varphi^\text{barrier}$; Thus, we
conclude that via a sudden quench of the barrier, and after a strong
measurement of the charge in the left side, one effectively obtains the
entangled state Eq.~(\ref{fusion}) for any outcome of the measured charge
$N_L$.

In
Appendix~\ref{sec:fusion_protocol_within_an_exactly_solvable_particle_number_conserving_model}
we complement this discussion of the fusion protocol in a number conserving
system by studying an exactly solvable toy model introduced by Iemini
\emph{et.~al.}~\cite{Iemini2015}. This allows for an explicit treatment of the
subtleties associated with the gapless charge sector.

\section{Conclusions}
In this work we described a closed, particle-conserving setup in a cold atom
system endowed with both a synthetic dimension and a synthetic gauge field with
the goal of realizing close analogues of Majorana zero modes. In contrast to
other approaches relying on the superconducting proximity effect, the
two-particle hopping process, which is responsible for the topological
superconducting phase, is generated in our setup from the single particle
hopping in the presence of small attractive interaction. We emphasize, that our
model builds upon an experimentally tested implementation of quantum Hall edge
states in the cold atom setting. 	

The synthetic dimension approach has a number of advantages. Local
particle-particle interactions in real space become non-local in the synthetic
dimension. This is the key feature which effectively turns a small local
attractive interaction into an attractive interaction between ``distant'' edge
states in a synthetic quantum Hall ribbon, thus resulting in the formation of
analogues of Majorana zero modes. The wavefunction of these modes, while local
in the real space, remains non-local in the synthetic dimension. The
prerequisite non-local interactions can not be induced in a conventional
condensed matter setup.

The fact that Majorana operators are non-local in the synthetic dimension
allows us to manipulate the associated degeneracy by adding a synthetic flux
$\Phi_x$. This in turn can provide us with a useful knob for combining
``topological'' and ``non-topological'' qubit operations. An even more
intriguing possibility is to utilize a similar setup for producing and
manipulating fractionalized zero
modes~\cite{Clarke2013a,Lindner2012,Cheng2012,Alicea2016}, a topic that we
intend to cover elsewhere.

\section{Acknowledgments}  	
The authors would like to thank Hans Peter B\"{u}chler, Sebastian Diehl, Yuval
Gefen, Leonardo Mazza, and Christophe Mora for useful discussion. This project
was supported in part by the ISF Grant No.~1243/13 and by the BSF Grant
No.~2016255; KS was supported in part by the NSF DMR-1411359 grant.

\bibliographystyle{apsrev4-1}

\appendix
\section{Matrix elements of $\mathcal{H}_\perp$}
Our goal is to compute matrix elements of $\mathcal{H}_\perp(x)$ in the ground state manifold.
To set up the notation, we write the mode expansion (in this appendix we set $\theta=\sqrt{K_-}\theta_-$, $\varphi=\varphi_-/\sqrt{K_-}$, $v_- = v$)
\bea
\varphi(t,x) =\varphi_0+ \sum_{k \ne 0} \frac{1}{\sqrt{4 \pi |n|}} \left( \beta_n e^{i k_n x - i v |k_n| t} + h.c.\right),
\eea
where $k_n=\frac{2 \pi n}{L}$  and $[\beta_n , \beta_{n'}]=\delta_{n,n'}$.
The free Hamiltonian is $H_0 = \frac{v}{2} \int_0^L dx [(\partial_x \varphi)^2+ \Pi^2]$ with the conjugate field $\Pi(t,x) = \frac{1}{v} \partial_t \varphi(t,x)$ satisfying $[\varphi(t,x) , \Pi(t,y)]=i \delta (x-y)$. We define also $\Pi(t,x) = - \partial_x \theta(x)$. One can check that these canonical commutation relations are recovered using the mode expansion and that the Hamiltonian becomes $H_0 =  \frac{2 \pi v}{L} \sum_{n=1}^{\infty} n [\beta_n^\dagger \beta_n+\beta_{-n}^\dagger \beta_{-n}]$.

We consider the interface $x=0$ between the region where $\theta$ is pinned
$\theta(x) = \sqrt{\pi/2}~n_\theta$ for $x>0$, and $\varphi$ is pinned to
$\varphi(x) = \sqrt{2 \pi} n_\varphi^L$ for $x<0$. The latter is accounted for
by the boundary condition $\varphi(x=0,t)=\sqrt{2 \pi} n_\varphi^L$. Using the
mode expansion, this boundary condition implies $\beta_n + \beta_{-n}=0$, hence
\bea \label{modeexpsincos}
\varphi(t,x) &=& \sqrt{2 \pi} n_\varphi^L+i \sum_{k \ne 0} \frac{1}{\sqrt{ \pi n}} \sin (k_n x) \left( \beta_n -\beta_n^\dagger \right), \nonumber \\
\theta(t,x) &=& \sum_{k \ne 0} \frac{1}{\sqrt{ \pi n}} \cos (k_n x) \left(
\beta_n +\beta_n^\dagger \right). \eea Ignoring tunneling between minima
$\theta(x) = \sqrt{\pi/2}~n_\theta$ of the cosine potential in the region
$x>0$, we replace the latter by a mass term $g (\theta-
\sqrt{\pi/2}~n_\theta)^2$ (where $g\rightarrow \tilde{g}=g/K_-$). This gives a
finite average value $\langle \theta(x) \rangle = \sqrt{\pi/2}~n_\theta$. To
account for fluctuations around this value, we rewrite the Hamiltonian using
the mode expansion: \be H_{n_\theta} = \sum_{n=1}^\infty [A_n (\beta^\dagger_n
\beta_n + \beta_n \beta_n^\dagger) + B_n (\beta_n^2+(\beta_n^\dagger)^2)], \ee
where $A_n = v k_n+ B_n$, $B_n = g/k_n$. After a Bogoliubov transformation one
obtains $H = \sum_n E_n b_n^\dagger b_n$ where $\beta_n = f_n  b_n + g_n
b_n^\dagger$, $f_n = \sqrt{\frac{1}{2} \left( \frac{A_n}{E_n} + 1 \right)}$,
$g_n  = - \sqrt{\frac{1}{2} \left( \frac{A_n}{E_n} - 1 \right)}$,
$\protect{E_n^2 =A_n^2 - B_n^2}$. The Hamiltonian $H_{n_\theta}$ describes
excitations above the $| n_\theta \rangle$ ground state.

Having argued that $\mathcal{H}_\perp$ is diagonal in this basis, we will
evaluate
\begin{multline}
\langle n_\theta |\mathcal{H}_\perp(x)| n_\theta \rangle
= t_\perp e^{i\Phi_x/2} \langle n_\theta |R^\dagger_{a}R_b | n_\theta \rangle\\
+
t_\perp e^{-i\Phi_x/2} \langle n_\theta |L^\dagger_{a}L_b| n_\theta +c.c.
\rangle
\end{multline}
This requires calculating quantities such as
\begin{equation*}
\langle n_\theta |R^\dagger_{a}R_b |
n_\theta \rangle
= \frac{1}{\pi a_0} \langle n_\theta |e^{- i
\sqrt{\frac{2 \pi}{K_-}} (\theta(x)+K_-\varphi(x))} | n_\theta \rangle .
\end{equation*}

Introducing a shorthand notation $ \langle ... \rangle\equiv \langle n_\theta|
...| n_\theta \rangle$ and using $\langle e^A \rangle = e^{\langle A \rangle}
e^{\frac{1}{2} \langle (A - \langle A \rangle )^2 \rangle}$, we have \be
\langle e^{i \sqrt{\frac{2 \pi}{K_-}} (\theta + K_-\varphi)} \rangle = e^{i \pi
n_\theta } e^{- \pi/K_- (\langle (K_-\varphi + \theta)^2  \rangle  -( \langle
K_-\varphi + \theta \rangle )^2 }. \ee
 Computing the fluctuations using Eq.~\eqref{modeexpsincos}
we obtain
\be
\langle e^{i \sqrt{\frac{2 \pi}{K_-}} (\theta (x)+ K_-\varphi(x) )} \rangle
=e^{i \pi n_\theta }    e^{- \pi {\text{var}} (\theta_-(x)) - \pi {\text{var}}
(\varphi_-(x))}, \ee and \bea {\text{var}} (\theta(x)) =
\sum_{n=1}^\infty \frac{\cos^2(k_n x)}{\pi n} \frac{v k_n}{\sqrt{(v k_n)^2+2 g v}}, \nonumber \\
{\text{var}} (\varphi(x)) =
\sum_{n=1}^\infty \frac{\sin^2(k_n x)}{\pi n} \frac{v k_n+\frac{2g}{k_n}}{\sqrt{(v k_n)^2+2 g v}}.
\eea
One can see a length scale $\xi = \sqrt{\frac{vK_-}{2g}}$. Evaluating these sums for $\xi \ll x \ll L$ gives
\bea
{\text{var}} (\theta_-(x)) &\sim &\frac{1}{2 \pi K_-} {\text {log}} (\xi/a_0), \nonumber \\
{\text{var}} (\varphi_-(x)) &\sim & \frac{K_-}{2 \pi} {\text {log}} (\xi/a_0)
+\frac{1}{2} \frac{xK_-}{\xi}. \eea
One can see then that
\begin{multline}
\langle n_\theta |R^\dagger_{a}R_b |
n_\theta \rangle = \langle n_\theta |L^\dagger_{a}L_b | n_\theta \rangle\\
=
\frac{e^{i \pi n_\theta }  e^{-\pi xK_-/(2
\xi)}a_0^{1/2K_-+K_-/2-1}}{\pi\xi^{1/2K_-+K_-/2}}
\end{multline}
The result of the calculation is then
\begin{multline}
\langle n_\theta |\mathcal{H}_\perp (x) | n_\theta
\rangle\\
= \frac{2 t_\perp e^{i \pi n_\theta } e^{-\pi xK_-/(2
\xi)}a_0^{1/2K_-+K_-/2-1}\cos (\Phi_x/2)}{\pi\xi^{1/2K_-+K_-/2}}.
\end{multline}
Integrating over $x$ one obtains
\begin{multline}
\langle n_\theta |H_\perp | n_\theta \rangle\\
= \frac{4 t_\perp}{\pi^2K_-} \cos (\Phi_x/2)   e^{i \pi n_\theta }
\left(\frac{a_0}{\xi}\right)^{1/2K_-+K_-/2-1}.
\end{multline}


\section{Fusion protocol within an exactly solvable particle-number conserving model} 
\label{sec:fusion_protocol_within_an_exactly_solvable_particle_number_conserving_model}
In order to better understand the physics of the fusion process in our model, we turn to an exactly-solvable toy model~\cite{Iemini2015}, which is particle-number conserving as well. We begin by reviewing the construction of the model~\cite{Iemini2015}, and then we describe the fusion process.

The well known Kitaev model~\cite{Kitaev2001} is a non particle-number conserving model whose Hamiltonian is
\begin{equation}
H_K = \sum\limits_j\left(-J a_j^\dagger a_{j+1} - \Delta a_j a_{j+1} +h.c. - \mu(n_j-\frac{1}{2}) \right).
\end{equation}
It is exactly solvable and at the ``sweet spot'' in the parameter space where
$\mu=0$, and $\Delta=J>0$ (we set $\Delta=J=1$), it is diagonalized as $H_K =
\sum_j \ell^\dagger_j\ell_j$, where $\ell_j = C^\dagger_j+A_j$, $C^\dagger_j =
a^\dagger_j+a^\dagger_{j+1}$, and $A_j = a_j-a_{j+1}$. In order to construct a
number conserving model out of the Kitaev model, we recall that in BCS theory
we begin with a number-conserving, quartic Hamiltonian, and then by employing a
mean field description, we turn into a quadratic Hamiltonian and lose number
conservation. Analogously, one can write the quartic Hamiltonian $H_a =
\sum\limits_j L^\dagger_j L_j$, where $L_j = C^\dagger_j A_j$. It is immediate
to see that this model is number conserving. Also the ground state of the
Kitaev model satisfying $\ell_j|\text{GS}_K\rangle = 0$, is the ground state of
this model as well since $L_j|\text{GS}_K\rangle =C_j^\dagger
A_j|\text{GS}_K\rangle  = C_j^\dagger C_j^\dagger|\text{GS}_K\rangle  = 0$. In
fact, one can project from the model states with well defined number of
particles, each of which is a ground state of $H_a$.

The next step is to create an interacting two-chain version of this model.
Denoting the chains by an extra subscript $a,b$, we do this by adding an
interaction term - $L_{I,j} = C^\dagger_{a,j}A_{b,j} + C^\dagger_{b,j}A_{a,j}$.
This interaction term is selected since it annihilates the Kitaev ground state
as well, while conserving the total particle number.

We can use this model to better understand the fusion process in our setup. The
fusion/splitting processes can be modeled by making coupling constants for the
central links $(\alpha=a,b)$ as well as central plaquette $(\alpha=I)$ time
dependent:
\begin{equation}
\hat{H} = \sum\limits_{\alpha=a,b,I}\sum\limits_{j\neq \ell/2}^{\ell-1} \hat{L}^\dagger_{\alpha,j}\hat{L}_{\alpha,j} + g(t)\sum\limits_{\alpha=a,b,I}\hat{L}^\dagger_{\alpha,\ell/2}\hat{L}_{\alpha,\ell/2},
\end{equation}
where $a,b$ label different chains and $\ell$ is the chain lenth (assumed to be
even). Initially $g(0)=1$, corresponding to a uniform system, and eventually
$g(t \to \infty) =0$, corresponding to two decoupled halves. It is important to
notice that each single term in this Hamiltonian annihilates the ground state.
Hence, if we begin in a ground state, the time dependent Schr\"odinger equation
trivially guarantees that we remain in the same ground state. Therefore for
this specific model we need not assume anything about the functional form of
the time dependence $g(t)$.

The ground states of this Hamiltonian at $g=1$ are characterized by the total
number of particles $N$, and by the parity of the two chains, being both even
or both odd for an even number of particles, or one even and one odd if $N$ is
odd. They are given by
\begin{equation}
\begin{gathered}
\label{eq:exact_gs}
|\psi_\ell(N)\rangle_{ee} = \mathcal{N}^{-1/2}_{ee,\ell,N}\sum\limits_{n=0}^{N/2}\sum\limits_{\substack{\{\vec{j}_{2n}\} \\ \{\vec{q}_{N-2n}\} }}|\vec{j}_{2n}\rangle \otimes |\vec{q}_{N-2n}\rangle, \\
|\psi_\ell(N)\rangle_{oo} = \mathcal{N}^{-1/2}_{oo,\ell,N}\sum\limits_{n=0}^{N/2-1}\sum\limits_{\substack{\{\vec{j}_{2n+1}\} \\ \{\vec{q}_{N-2n-1}\} }}|\vec{j}_{2n+1}\rangle \otimes |\vec{q}_{N-2n-1}\rangle.
\end{gathered}
\end{equation}
where $\vec{j}_m,\vec{q}_m$ are m-sized vectors denoting the ordered locations of particles on the $a$ and $b$ chains respectively, and the sum is over all possible configurations,
\bea
\mathcal{N}_{ee,\ell,N}&=& \sum_{n=0}^{N/2}\binom{\ell}{2n}\binom{\ell}{N-2n},~~(N~{\rm{even}})  \nonumber \\
\mathcal{N}_{oo,\ell,N}&=& \sum_{n=0}^{N/2-1}\binom{\ell}{2n+1}\binom{\ell}{N-2n-1}.
\eea
Similarly for $N$ odd the number of configurations with well defined parities in each chain is
\bea
\mathcal{N}_{eo,\ell,N}=\mathcal{N}_{oe,\ell,N}= \sum_{n=0}^{(N-1)/2}\binom{\ell}{2n}\binom{\ell}{N-2n}. ~~(N~{\rm{odd}}) \nonumber
\eea

In the end of the fusion process $g=0$ and the system separates into two
halves. Since the wave function in this model does not change at all, in order
to describe the final state we simply perform a left-right decomposition of the
ground state wave function. We assume that the total number of particles $N$ as
well as well as the number of particles in each individual chain is even. Since
the system remains in the zero-energy ground state of the decoupled left-right
Hamiltonians, it is guaranteed that we can use the same ground states
Eq.~(\ref{eq:exact_gs}) of the two half-systems $\ell \to \ell /2$. We find
that the decomposition is given by
\begin{multline}
|\psi_\ell(N)\rangle_{ee} = \sum\limits_{\mu=e,o} \sum\limits_{\substack{N_L=0 }}^N A_{N_L,\mu\mu'}
|\psi^L_{\ell/2}(N_L)\rangle_{\mu\mu'}\\
\otimes|\psi^R_{\ell/2}(N-N_L)\rangle_{\mu\mu'},
\end{multline}
Here $N_L$ is the charge of the left part of the system. Writing \be N= n_a
+n_b,~N_L= n_a^L +n_b^L,~~~N-N_L= n_a^R +n_b^R, \ee since we assumed $n_a$ and
$n_b$ are even, the parity of $n_a^L$ equals that of $n_a^R$ and is denoted
$\mu$; then the equal parities of $n_b^L$ and $n_b^R$ which are denoted $\mu'$
equal \be \mu'=\mu ({\rm{for}}~N_L~{\rm{even}}),~~~
\mu'=\bar{\mu}~{\rm{for}}~N_L~{\rm{odd}}, \ee with $\bar{o}=e$ and $\bar{e}=o$.
The coefficients are given by $A_{N_L,\mu \mu'} =
\sqrt{\frac{\mathcal{N}_{\mu\mu',\ell/2,N_L}\mathcal{N}_{\mu\mu',\ell/2,N-N_L}}{\mathcal{N}_{ee,\ell,N}}}$.
In fact the dependence of these coefficients on $\mu , \mu'$ becomes
exponentially small with increasing system size. Ignoring these exponentially
small effects we have $A_{N_L,\mu \mu'}=A_{N_L}$ and can therefore write
\begin{multline}
|\psi_\ell(N)\rangle_{ee}\\ =  \sum\limits_{\substack{N_L=\text{even} }} A_{N_L}
\left[|\psi^L_{\ell/2}(N_L)\rangle_{ee} \otimes|\psi^R_{\ell/2}(N-N_L)\rangle_{ee}\right.
  \\
+\left. |\psi^L_{\ell/2}(N_L)\rangle_{oo} \otimes|\psi^R_{\ell/2}(N-N_L)\rangle_{oo}  \right]  \\
+ \sum\limits_{\substack{N_L=\text{odd} }} A_{N_L} \left[|\psi^L_{\ell/2}(N_L)\rangle_{eo} \otimes|\psi^R_{\ell/2}(N-N_L)\rangle_{eo}\right.
 \\
+ \left.|\psi^L_{\ell/2}(N_L)\rangle_{oe} \otimes|\psi^R_{\ell/2}(N-N_L)\rangle_{oe}  \right]
\end{multline}
We can now explicitly see that for each particle number $N_L$ in the left
segment, the parity of each individual chain is not well defined and is in fact
entangled between the left and right sides in accordance to the fusion rules
for Majorana zero modes $\gamma_3,\gamma_4$.

However one should note a peculiarity of this model. In this equation all
charge states are equally likely. The degeneracy of the system is extensive as
all these states differing by $N_L$ are degenerate, in addition to the
degeneracy associated with the parity of each chain in each side, attributed to
Majorana operators. In any generic system like the one considered in the main
text, these states differing by $N_L$ are non-degenerate and actually have an
energy difference which scales with the system size as power law. This can be
understood using bosonization. While the odd sector is gapped everywhere,
either by pair tunneling in the superconductor regions or by the mass term in
the trivial regions, the even sector, i.e. the charge sector, is only gapped in
the trivial regions. The superconductor regions act as Luttinger liquids for
the charge sector. Hence they provide a finite stiffness, which favors a state
with a fixed charge.

\end{document}